\documentclass[12pt]{iopart}

\expandafter\let\csname equation*\endcsname=\relax 
\expandafter\let\csname endequation*\endcsname=\relax 

\usepackage{caption}
\usepackage[dvipsnames]{xcolor}
\usepackage{graphicx}
\usepackage{tabularx}
\usepackage{amssymb}
\usepackage{multirow}
\usepackage{braket}
\usepackage{booktabs}    

\usepackage{listings}
\usepackage[frozencache,cachedir=.]{minted}
\usepackage{chemformula}
\usepackage{textcomp}
\newcommand{\op}[1]{\hat {#1}}
\newcommand{\dm}{\op{\rho}}
\newcommand{\trace}[1]{\mathrm{Tr}\left(#1\right)}
\newfloat{code}{tbp}{loc} 
\floatname{code}{Listing}
\setminted{fontsize=\footnotesize}\usepackage[most]{tcolorbox}

\newcommand{\remotemanager}{{\lstinline{remotemanager}}}

\lstdefinestyle{python-style}{
  breaklines=true,
  frame=L,
  commentstyle=\color{purple!40!black},
  identifierstyle=\color{black},
  stringstyle=\color{blue},
  language=Python,
  captionpos=b,
  basicstyle=\footnotesize
}

\begin{document}

\newcommand{\WD}[1]{\textcolor{purple}{[WD: #1]}}
\newcommand{\LR}[1]{\textcolor{blue}{[LR: #1]}}

\title[]{Exploratory Data Science on Supercomputers for Quantum Mechanical Calculations}

\author{William Dawson}
\address{RIKEN Center for Computational Science, Kobe, Hyogo, 650-0047, Japan}
\author{Louis Beal}
\address{Univ.\ Grenoble Alpes, MEM, L\_Sim, F-38000 Grenoble, France}
\author{Laura E. Ratcliff}
\address{Centre for Computational Chemistry,
School of Chemistry, University of Bristol, Bristol BS8 1TS,
United Kingdom}
\address{Hylleraas Centre for Quantum Molecular Sciences, Department of Chemistry, UiT The Arctic University of Norway, N-9037 Troms\o{}, Norway}
\author{Martina Stella}
\address{The Abdus Salam International Centre for Theoretical Physics (ICTP), Condensed Matter and Statistical Physics, 34151 Trieste, Italy}
\author{Takahito Nakajima}
\address{RIKEN Center for Computational Science, Kobe, Hyogo, 650-0047, Japan}
\author{Luigi Genovese$^\dagger$}
\address{Univ.\ Grenoble Alpes, MEM, L\_Sim, F-38000 Grenoble, France}
\ead{luigi.genovese@cea.fr}

\vspace{10pt}
\begin{indented}
\item[]October 2023
\end{indented}

\begin{abstract}
Literate programming --- the bringing together of program code and natural language narratives --- has become a ubiquitous approach in the realm of data science. This methodology is appealing as well for the domain of Density Functional Theory (DFT) calculations, particularly for interactively developing new methodologies and workflows. However, effective use of literate programming is hampered by old programming paradigms and the difficulties associated with using High Performance Computing (HPC) resources. Here we present two Python libraries that aim to remove these hurdles. First, we describe the PyBigDFT library, which can be used to setup materials or molecular systems and provides high-level access to the wavelet based BigDFT code. We then present the related \remotemanager{} library, which is able to serialize and execute arbitrary Python functions on remote supercomputers. We show how together these libraries enable transparent access to HPC based DFT calculations and can serve as building blocks for rapid prototyping and data exploration.
\end{abstract}

\section{Introduction}

In the field of quantum mechanical (QM) modeling, popular theories such as Density Functional Theory (DFT)~\cite{hohenberg-inhomogeneous-1964, kohn-self_consistent-1965} have the ability to predict the properties of both solid-state and molecular systems. While the conventional wisdom has been that DFT cannot blindly be applied as a black box method, theoretical progress and the development of best practices~\cite{Bursch2022} have made it appealing for use as a building block for new advances and workflows. This type of use case is particularly potent at the intersection of computational chemistry and data science. At the same time, advances in algorithms and high performance computing (HPC) are enabling the application of DFT to larger and larger systems, making it a tool of interest for new disciplines~\cite{Dawson2021}.

One example of using DFT as a building block is for the development of machine-learning models~\cite{Schmidt2019}. This leads to requirements such as robust calculations, high-throughput runs, and careful data providence. Another example is the encapsulation of DFT based workflows for computing complex properties~\cite{Curtarolo2012,Mathew2017,UHRIN2021110086}. For these purposes, the DFT part is only a small, well-defined kernel, often called from a high-level scripting language and sitting in a much larger framework. In principle, scientists can develop these kinds of frameworks in any of the many open source DFT packages that are available today~\cite{Lehtola2022}; however, old programming models and data structures with a narrow scope make this challenging. Several more recent electronic structure codes have tried to address this through alternative programming paradigms such as using modern C++~\cite{Peng2020, Kowalski2021}, Python~\cite{Enkovaara2010,Sun2020,Turney2012,Rinkevicius2020}, or Julia~\cite{Poole2020, Fathurrahman2020, Aroeira2022, DFTKjcon}. 

In the area of data science, literate programming~\cite{Knuth1984}, such as through Juypter virtual laboratory notebooks, has become ubiquitous~\cite{10.1145/3173574.3173748}. In this approach code, visualizations, and natural language narratives are intermingled into one unit. This practice has similar requirements to the previous areas in that DFT calculations should be based on high-level scripting and well-defined data structures. On the other hand, this kind of work often emphasizes exploratory calculations and short workflows that bring together methodologies and results~\cite{10.1145/3512934}. Virtual notebooks are often used for interactive programming, where early calculation results drive further code development --- a ``write-eval-think-loop'' as described by Granger and P{\'e}rez~\cite{Granger2021}. Thus, features such as databases, complex dependencies, and robust calculations become less important than development speed and legibility.

In this Technical Note, we will describe two Python based libraries we have developed for use with literate programming: PyBigDFT and \remotemanager. The first package, PyBigDFT, serves as a high level interface to set up, run, and post-process DFT calculations, specifically using the BigDFT code~\cite{Ratcliff2020}. The \remotemanager{} package is a general library for performing remote procedure call using Python on supercomputers. We will first provide overviews of the two libraries, followed by comparisons to existing approaches. Then we will present some sample applications where these two libraries together can enable new and complex calculation workflows. For all of the workflows presented here, we have included matching Jupyter notebooks on Github (github.com/BigDFT-group/resources). 

\section{PyBigDFT and \remotemanager}

Two features of BigDFT have served as core principles for the development of the Python libraries presented here. The first feature is the use of YAML as a language to write input and output files. YAML allows for the serialization of complex data structures by composing simple data types in a hierarchical fashion. YAML can thus serve as an intermediary between the ecosystems of a Python and Fortran code using dictionary data types (built in for Python, from our Futile library for Fortran). The second feature is the use of optional calculations in BigDFT. When a calculation is run, users can pass a flag to check if a previous calculation with that same name had completed successfully; if so, the calculation process returns immediately. This combination of serializability and lazy evaluation is manifest in both the overall PyBigDFT library and the \remotemanager{} facility. Together, they can be used to build virtual lab notebooks that enable new and complex workflows. 

\subsection{System Manipulation and Calculation with PyBigDFT}

Broadly, PyBigDFT can be divided into two sets of capabilities: (1) data structures for manipulating atomic systems and (2) functionality for driving calculations. The need for system manipulation may seem unintuitive to a user coming from codes that mainly represent the structures of interest as simple text files (XYZ, POSCAR, etc). However, large systems often have a hierarchical structure, and users may wish to exploit this structure for manipulation. Such complicated systems can subsequently be decorated with the results of DFT calculations.

In the spirit of the YAML approach, PyBigDFT hierarchically describes a system by composing simple data types (List.~\ref{lst:system}). At the lowest level is the \lstinline{Atom} class, which is a named collection of basic data types. This flexible data structure allows users to store arbitrary data with a given \lstinline{Atom}, which is useful both for pre-processing a system (storing of atom names, sources, etc), and when extracting the results of a BigDFT calculation (forces, charges, multipoles, etc). These atoms are organized into a list called a \lstinline{Fragment}. A \lstinline{Fragment} may be a layer of a solid-state system, a residue of a protein, a moiety of a molecular crystal, etc. These \lstinline{Fragment} objects are then further organized into a named collection called a \lstinline{System}. In principle, any hashable value may be used for a fragment name, but we have adopted the convention of \lstinline{NAME:ID} as inspired by the residue naming scheme of proteins. This hierarchical arrangement means that the atoms are in principle not ordered. The \lstinline{System} class provides a crucial function called \lstinline{compute_matching}, which can efficiently generate a map between a \lstinline{System} and an arbitrary set of \lstinline{Atom} objects so that data may be passed to and from programs that don't use a hierarchical representation or respect atom order.  PyBigDFT also has support for extended systems through its \lstinline{UnitCell} class. As the \lstinline{Fragment} and \lstinline{System} class only store shallow copies of their underling data, multiple views of the object of study (\lstinline{System} classes) can be employed.

\begin{code}
\begin{minted}{python}
# Create Three Atoms
at1 = Atom({"r": [0, 0, 0], "sym": "H"})
at2 = Atom({"r": [0, 0, 1.4], "sym": "H"})
at3 = Atom({"r": [10, 0, 0], "sym": "He"})

# Construct a System from Two Fragments (H2, He)
sys = System()
sys["H2:1"] = Fragment([at1, at2])
sys["He:2"] = Fragment([at3])

# Iterate Over The System
for fragid, frag in sys.items():
    for at in frag:
        print(fragid, at.sym, at.get_position())
\end{minted}
\caption{The hierarchy of \lstinline{Atom}, \lstinline{Fragment}, and \lstinline{System} in PyBigDFT. Three \lstinline{Atom} types are created, which are divided into two \lstinline{Fragment} objects, and then combined into one \lstinline{System} that can be accessed like a standard Python dictionary.}
\label{lst:system}
\end{code}

To perform a calculation, users must construct an \lstinline{Inputfile} and a \lstinline{SystemCalculator} (List.~\ref{lst:calc}). BigDFT, like many electronic structure codes, has numerous input parameters that control the various named modules. This motivates the definition of an input file as a hierarchy of dictionaries. When the \lstinline{SystemCalculator} is run, the \lstinline{System} and \lstinline{Inputfile} are serialized as YAML that is then processed by BigDFT. The optional \lstinline{skip} parameter can be used to trigger lazy evaluation. The result of a calculation is a \lstinline{Logfile}, which is generated from the YAML output of BigDFT. Users can access the full hierarchy of data present in the \lstinline{Logfile} using dictionary syntax. 

\begin{code}
\begin{minted}{python}
# Create an Input File
inp = Inputfile()
inp.set_hgrid(0.4)
inp.set_xc("PBE")
inp["perf"] = {"calculate_forces": False,
               "multipole_preserving": True}

# Generate a Calculator and Run
calc = SystemCalculator(skip=True)
log = calc.run(sys=sys, input=inp, 
               name="quick", run_dir="scratch")

# Access the Generated Logfile
print(log.energy)
print(log.log["Memory Consumption Report"]
             ["Memory occupation"])
\end{minted}
\caption{A BigDFT calculation driven by PyBigDFT. First an \lstinline{Inputfile} and \lstinline{SystemCalculator} are created. The result of a run is stored in a \lstinline{Logfile}.}
\label{lst:calc}
\end{code}

PyBigDFT also provides support for reading and writing standard atomic structure file formats, computing and plotting the Density of States, and visualization of molecules with their computed properties. To supplement the features of PyBigDFT, we provide interoperability modules to other popular packages. For system manipulation, we provide converters to and from ASE~\cite{HjorthLarsen2017}, OpenBabel~\cite{OBoyle2011}, and RDKit~\cite{rdkitRDKit}. For calculations using different quantum mechanical methods, we provide analogous classes to \lstinline{SystemCalculator} and \lstinline{Logfile} for DFTB+~\cite{Hourahine2020}, MRChem~\cite{Wind2023}, PSI4~\cite{Turney2012}, and XTB~\cite{Bannwarth2021}.

\subsection{Flexible Access to HPC Resources with \remotemanager}

A significant weakness of the initial PyBigDFT implementation was that the writing of workflows would be interrupted by the need to run computationally heavy calculations on remote machines. If a workflow was being written in a Jupyter notebook, one would need to convert that notebook to a script, submit it through the job system, wait for the calculation to complete, copy data to the machine with the Jupyter server, and then restart the notebook (exploiting lazy calculations). Computationally demanding calculations include not just formulaic BigDFT runs, but a wide variety of pre- and post-processing operations. These requirements necessitate a strategy for driving the execution of \emph{arbitrary} Python routines on (potentially multiple) remote machines.

To fulfill this requirement we have developed a new library called \remotemanager{} (List.~\ref{lst:remote}). This library is able to serialize arbitrary Python routines and arguments that are sent to a remote machine and executed through a job system, after which the resulting data are synchronized to the local machine. For simple data types, serialization is done with YAML. For more complex data types, external libraries like Dill or JSONPickle can be selected. To use \remotemanager{}, a user defines the function they want to run remotely, and uses it to create a \lstinline{Dataset}. Sets of arguments are then added to the \lstinline{Dataset} and the whole set can be run asynchronously. If auxiliary functions are required, they can be defined with the \lstinline{@RemoteFunction} decorator. The \remotemanager{} package follows the model of BigDFT's \lstinline{SystemCalculator} in that a \lstinline{skip} argument will check if the calculation has already completed, and if so no calculation is performed. 

\begin{code}
\begin{minted}{python}
# Function to Run on the Remote Machine
def remote_function(sys, inp, run_name):
    from BigDFT.Calculators import SystemCalculator
    calc = SystemCalculator()
    log = calc.run(sys=sys, input=inp, name=run_name)
    return log.energy

# Create a Dataset to Store the Function
remote = Dataset(function = remote_function,
                 url = url,
                 serialiser = serialdill())

# Append a Run with Data
args = {"sys": sys, "inp": inp, "run_name": "remote"}
remote.append_run(arguments = args)

remote.run()  # Submit the Calculation

# Wait for Job Completion, Checking Every 10 Seconds
remote.wait(interval=10, timeout=3600)

# Fetch and Display the Results
remote.fetch_results()
print(f'energy: {remote.results}')
\end{minted}
\caption{An example of a remotely submitted calculation. First we define a function that we wish to run remotely. The function is given to the \lstinline{Dataset} object along with a \lstinline{URL} object that describes the remote computer. The arguments to the function are given to the \lstinline{append_run} method. Finally, the run method serializes everything, submits a calculation asynchronously, waits for the calculation to complete, and fetches the result.}
\label{lst:remote}
\end{code}

The \remotemanager{} library does not use a daemon on the remote machine and only needs to be installed on the user's workstation. Collections of remote calls can be bundled together into the \lstinline{Dataset} type, which is optimized to minimize commands sent to the remote machine (a feature that helps in high latency environments). Connections are defined (List.~\ref{lst:connection}) via the \lstinline{URL}  module: a Python wrapper around the Secure Shell Protocol (SSH). On top of the \lstinline{URL} class is the \lstinline{Computer} derived type and associated derived types for various specific machines. Each computer type is defined with information about how to generate job scripts and the associated options (number of nodes, maximum wall time, etc). Front end versions of a machine can also be defined for low cost operations. A user planning to access a new machine can either define a new \lstinline{Computer} derived type or define one using a YAML configure file.

\begin{code}
\begin{minted}{python}
# Basic Url
url = URL(user="username", host="remote.host.address")
url.test_connection()

# YAML Configured
url = BaseComputer.from_yaml("summer.yaml", user="username")

# Or Derived Type Computer
url = Fugaku(user="username")

# Computer Configuration
url.mpi = 48
url.omp = 12
url.nodeshape = "torus"  # Machine Specific
\end{minted}
\caption{Definition of the \lstinline{URL} base class and predefined \lstinline{Computer} types. A \lstinline{URL} can be defined as a simple remote address, with a configuration file, or as a derived type when complex logic is required for jobscript generation and submission. After a \lstinline{URL} object is constructed, global job parameters can be set.}
\label{lst:connection}
\end{code}

\subsection{Literate Programming and Computational Continuity}

Code development inside virtual notebooks often differs substantially from best practices in software engineering~\cite{10.1145/3524842.3528447, 10.1145/3512934}. As noted above, developers frequently use notebooks for exploratory calculations instead of building well structured workflows. The \remotemanager{} library can work well under these circumstances (List.~\ref{lst:magic}). Entire Jupyter cells can be designated to run remotely using the ``cell magics'' included with \remotemanager{} (keyword \lstinline{sanzu}). Thus, we envision \remotemanager{} as a general library for performing Remote Procedure Call on HPC systems, which enables explorations that may have large computational requirements. The intermingling of text and code in literate programming make notebooks a perfect tool for collaborative efforts. The \remotemanager{} framework can enable such collaborative use and ``Computational Continuity'': allowing users to hand off notebooks and serialized results to their coworkers who can then add to the notebook and generate further results without having to redo computationally demanding portions.

\begin{code}
\begin{minted}{python}
# [Cell 1: Run Dataset - Computing Energy of Many Systems]
for name, sys in systems.items():
    args = {"sys": sys, "inp": inp, "run_name": name}
    remote.append_run(arguments = args)
remote.run(); remote.wait() ; remote.fetch_results()

# [Cell 2: Process The Results - Find The Lowest Energy System]
energies = {k: v for k, v in zip(systems, remote.results)}
mink = min(energies, key=energies.get)
minsys = systems[mink]

# [Cell 3: Magic Single Job - Restart the Job and Write Cubefiles]
%%sanzu url=url, serialiser=sdill
%%sargs sys=minsys, inp=inp, run_name=mink
from BigDFT.Calculators import SystemCalculator
inp.read_orbitals_from_disk()
inp.write_cubefiles_around_fermi_level()
calc = SystemCalculator()
log = calc.run(sys=sys, input=inp, name=run_name)
\end{minted}
\caption{Jupyter cell executed remotely with cell magics (\lstinline{sanzu}). After a set of systems is processed, a single post-processing operation is defined, which can be run without the \lstinline{Dataset} boilerplate. The \lstinline{sargs} delimiter is used to indicate arguments to the function executed remotely. Return values, printed data, and errors appear in the notebook as if the cell was run locally.}
\label{lst:magic}
\end{code}

\subsection{Handling Failure}

In exploratory calculations, programming mistakes are common. Developers will write an erroneous function to run remotely, arguments maybe be specified incorrectly, the remote environment could be setup wrong, job submission can fail, and a calculation might require more wall time than requested. As \lstinline{remotemanager} is focused on exploratory workflows, it is essential to provide troubleshooting capabilities  (List.~\ref{lst:fail}). In \remotemanager{}, runs are considered ``complete'' even if they fail; a summary of the captured error message is available in the \lstinline{errors} member variable of the \lstinline{Dataset}. When the \lstinline{Dataset} constructor is called with a modified function, a fresh \lstinline{Dataset} is automatically created (by comparing the function hash to the existing \lstinline{Dataset}) and the old \lstinline{Dataset} stored to a backup file; a subsequent run command will cause all calculations to be performed from scratch. If the error is caused by the remote environment, and a \lstinline{Dataset} should be rerun without changing the function, the state of a \lstinline{Dataset} can be reset by calling \lstinline{wipe_runs}.

\begin{code}
\begin{minted}{python}
# Broken Function
def f(x):
    returnnext(open(str(x**2)))
ds = Dataset(f)
ds.append_run({"x": "3"}); ds.run(); ds.fetch_results()
## NameError: name 'returnnext' is not defined
print(ds.errors[0])

# Fixed Function, Broken Argument
def f(x):
    return next(open(str(x**2)))
ds = Dataset(f)
ds.append_run({"x": "3"}); ds.run(); ds.fetch_results()
## TypeError: unsupported operand type(s) for ** or pow(): 'str' and 'int'
print(ds.errors[0])

# Fixed Argument
ds.remove_run(0)
ds.append_run({"x": 3}); ds.run(); ds.fetch_results()
## FileNotFoundError: [Errno 2] No such file or directory: '9'
print(ds.errors[0])

# Fixed Environment
open("temp_runner_remote/9", 'w').write("data")
ds.runners[0].run(force=True); ds.fetch_results()
## None
print(ds.errors[0])
\end{minted}
\caption{The \lstinline{remotemanager} library provides  access to Python error messages in failed runs so that remote functions can be debugged. By modifying the function or manipulating individual runners calculations can be resubmitted.}
\label{lst:fail}
\end{code}

A \lstinline{Dataset} contains a list of individual \lstinline{Runner} objects, which allow for fine grained control. Each \lstinline{Runner} object contains the state of a given run, such as whether it is finished, was successful, or had errors or warnings. When the arguments of a given run are incorrectly specified, the associated \lstinline{Runner} can be manually removed from the \lstinline{Dataset} with the \lstinline{remove_run} function. If, for example, a given runner needed more wall time or failed due to a missing file, that run can be resubmitted with the \lstinline{force=True} option. The \remotemanager{} library can thus be used to interactively build workflows that run on remote machines using a trial and error processes with limited cognitive overhead.

\section{Comparison With Other Packages}

Numerous packages exist that provide high level access to quantum chemistry programs. The solutions available attempt to address various subsets of requirements: ease of use, ease of development, addressing complex systems, encapsulation of common workflows, orchestration of multiple tools, data providence, and the need for high-throughput calculations. Here we will outline some of those packages and compare them to the PyBigDFT and \remotemanager{} combination. 

\subsection{Structures and Calculations}

There is significant interest in creating high level interfaces to existing QM codes, including the use of Graphical User Interfaces (GUI) and high level programming languages~\cite{Sherrill2020}. Web based interfaces also promise to offer easy access and ease of use (see, for example, CalcUS~\cite{Robidas2022} and the references therein). The focus of our developments is to empower capable programmers to be able to better exploit QM codes as building blocks, as opposed to simplifying access for novices (as a GUI would). 

As PyBigDFT exposes core data structures for manipulating molecular systems, it can be viewed as similar to tools like OpenBabel~\cite{OBoyle2011} or RDKit~\cite{rdkitRDKit}. These libraries offer a wider range of tools than PyBigDFT's \lstinline{System} class, which is why we have built interoperability modules between them. The PyBigDFT representation remains ideal to manipulate the system at various levels of a hierarchy.  Inside the QMflows~\cite{Zapata2019} package is the Molkit library for manipulating structures. It also includes the PLAMS library of Amsterdam Density Functional (ADF)~\cite{teVelde2001} that has a similar structure to PyBigDFT's \lstinline{Inputfile} and the ability to run and manage remote jobs.

Our approach also differs from codes like GPAW~\cite{Enkovaara2010}, PSI4~\cite{Turney2012}, or PySCF~\cite{Sun2020} that aim to fully expose the underlying QM data structures in Python. These packages are extremely useful for developing methods, but come at the cost of reengineering a code from scratch and greater challenges when targeting exotic HPC platforms. In our approach, the only requirements for the compute nodes is to be able to compile the Fortran90 BigDFT program and run the Python (version 3.5+) boilerplate code. This makes it similar to the client-server mode available in the Qbox code~\cite{Gygi2008}, with the added benefit of the full Python ecosystem. TeraChem also has a client server mode based on Google's Protocol Buffers that is effectively used for the TeraChem Cloud framework~\cite{Seritan2020}. Our approach is perhaps most similar to the recent Dalton Project and its associated PyFRAME package for multiscale modeling~\cite{Olsen2020}. Dalton Project shares many of the same design considerations: the existence of a mature code base and large systems as a target. 

A number of high level platforms exist that are able to target multiple electronic structure codes. The Atomic Simulation Environment (ASE)~\cite{HjorthLarsen2017} supports dozens of different codes, has tools for building systems (especially materials), and modules for post-processing calculations. The Atomic Simulation Recipes extension~\cite{Gjerding2021} builds on ASE to provide a number of recipes for common simulation tasks and can be run remotely through its MyQueue module. PyBigDFT is similar in that it has data structures to manage systems and calculator classes, and is used as glue code to build a BigDFT ASE calculator. ChemShell~\cite{Metz2014,Lu2019} focuses on QM/MM calculations, and thus can setup multiscale systems, run calculations using multiple codes, and includes a task farming capability for HPC resources. 

\subsection{Remote Execution}

The \remotemanager{} package on its own might be compared to industry tools aimed at cloud computing resources. It essentially implements remote procedure call, which is available in libraries like RPyC. In our case, we provide a middle layer that handles the concept of job submission ubiquitous on HPC resources. A number of tools have been developed to access cloud computing resources such as Hadoop or Apache-Spark. The Dask library includes dynamic task scheduling for building complex workflows. Parsl~\cite{10.1145/3307681.3325400} has been designed to run task based Python programs on HPC machines.

An important recent trend is the use of high-throughput QM calculations. The fireworks framework is able to orchestrate complex workflows of scripts~\cite{CPE:CPE3505}, and has been used extensively for high-throughput studies. Atomate~\cite{Mathew2017}, which builds off of fireworks and pymatgen~\cite{Ong2013}, provides templates for common materials workflows. The AFLOW~\cite{Curtarolo2012} library can be used for high-throughput calculations and has been used to build up a database of millions of materials. The combination of QM calculations and workflows on HPC resources makes the PyBigDFT - \remotemanager{} combination similar to AiiDA~\cite{Huber2020,UHRIN2021110086,Yakutovich2021}. In AiiDA there is a strong emphasis on reproducibility and persistence of data and workflows. The implementation and setup of \remotemanager{} is significantly simpler: it uses YAML files rather than the SQL database of AiiDA's persistence layer and is built with basic commands like rsync and SSH instead of a dedicated daemon. For well-defined projects with a focus on the resulting data, AiiDA may be the tool of choice, whereas \remotemanager{} is a powerful solution for exploratory calculations. The pyiron~\cite{pyiron-paper}  library also provides high-level access to QM programs, alongside utilities to set up systems, run jobs, and store calculation results in a database.

\section{Complex Workflows For Production}

Here we present some sample uses of the BigDFT and \remotemanager{} combination. We emphasize that these are not predefined workflows with well-defined input parameters. Rather, they are custom combinations of calculation and analysis built on high-level Python data structures. In addition to the full Jupyter notebooks available on Github, we also present program snippets that demonstrate key operations.

\subsection{Benchmarking}

The \remotemanager{} package can significantly simplify the process of benchmarking new computers or calculation methods. Benchmarking a new platform aligns well with literate programming: users often first carry out exploratory calculations on small partitions that lead to know-how and rules of thumb that can be embedded in natural language stories and later code that is run on a large number of nodes. 

\begin{code}
\begin{minted}{python}
%%sanzu url=furl, remote_dir=remote_dir
%%sanzu dbfile="scale_fetch_" + computer
%%sargs geoms=geoms, mpi=large_mpi, omp=large_omp
# Gather Timing Statistics
timing = {}
for g in geoms:
    run_name = g + "_" + str(mpi) + "_" + str(omp)
    with open("time-" + run_name + ".yaml") as ifile:
        full = load(ifile, Loader=SafeLoader)
    timing[run_name] = {"time": full["WFN_OPT"]["Classes"]["Total"][1]}

# Gather Memory Statistics
memory = {}
for g in geoms:
    run_name = g + "_" + str(mpi) + "_" + str(omp)
    with open("log-" + run_name + ".yaml") as ifile:
        full = load(ifile, Loader=SafeLoader)
    memory[run_name] = full["Memory Consumption Report"]\
                           ["Memory occupation"]["Peak Value (MB)"]

timing, memory # Final Values Returned
\end{minted}
\caption{Jupyter magic cell used to extract timing data and peak memory use from the BigDFT logfiles. This remote function is run once and synchronously, making it ideal for the ``cell magics'' approach. If additional properties are needed, any modification of the cell will cause it to run again from scratch.}
\label{lst:bmag}
\end{code}

Starting with a cluster of 2CzPN (1,2-bis(carbazol-9-yl)-4,5-dicyanobenzene) containing 1000 molecules (54000 atoms)~\cite{10.1103/PhysRevMaterials.1.075602}, we use PyBigDFT to extract subsets of various sizes. The benchmarking notebook then generates and launches jobs on the target machine with the desired number of nodes and threads using a \lstinline{Dataset}. After the \lstinline{Dataset} is computed, a magic cell is employed to run a Python function on the front end of the supercomputer that extracts the timing and memory statistics from each calculation and returns it as a standard Python dictionary (List.~\ref{lst:bmag}).

\begin{figure}
    \centering
    \includegraphics[width=1\columnwidth]{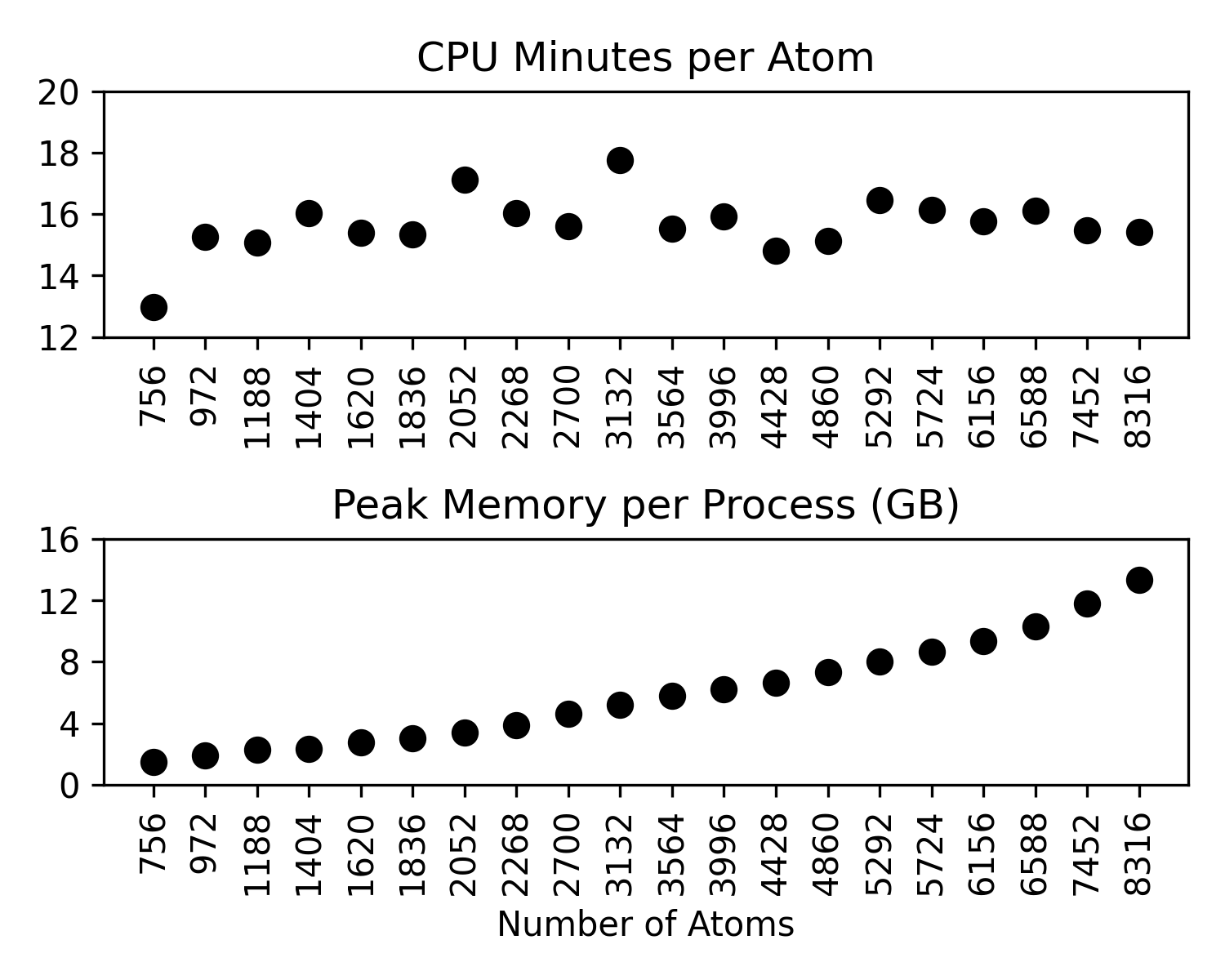}
    \caption{Calculation time and peak memory consumption (per process) of calculations of 2CzPN clusters of increasing size.}
    \label{fig:benchmark}
\end{figure}

As an example, we show the performance of BigDFT on the HPE Apollo2000 Gen10 Plus supercomputer located at the Research Center for Computational Science in Okazaki, Japan. Each node has two AMD EPYC 7763 processes with a total of 128 cores. We used 8 nodes for each calculation, with 16 MPI tasks and 8 OpenMP threads (a configuration informed by the single node exploratory calculations in the notebook). The CPU time per atom is reported in Figure~\ref{fig:benchmark}. We observe that we quickly reach the linear scaling regime even with around 1000 atoms. The combination of PyBigDFT and \remotemanager{} make it relatively easy to configure and benchmark new machines.

\begin{figure}
    \centering
    \includegraphics[width=1\columnwidth]{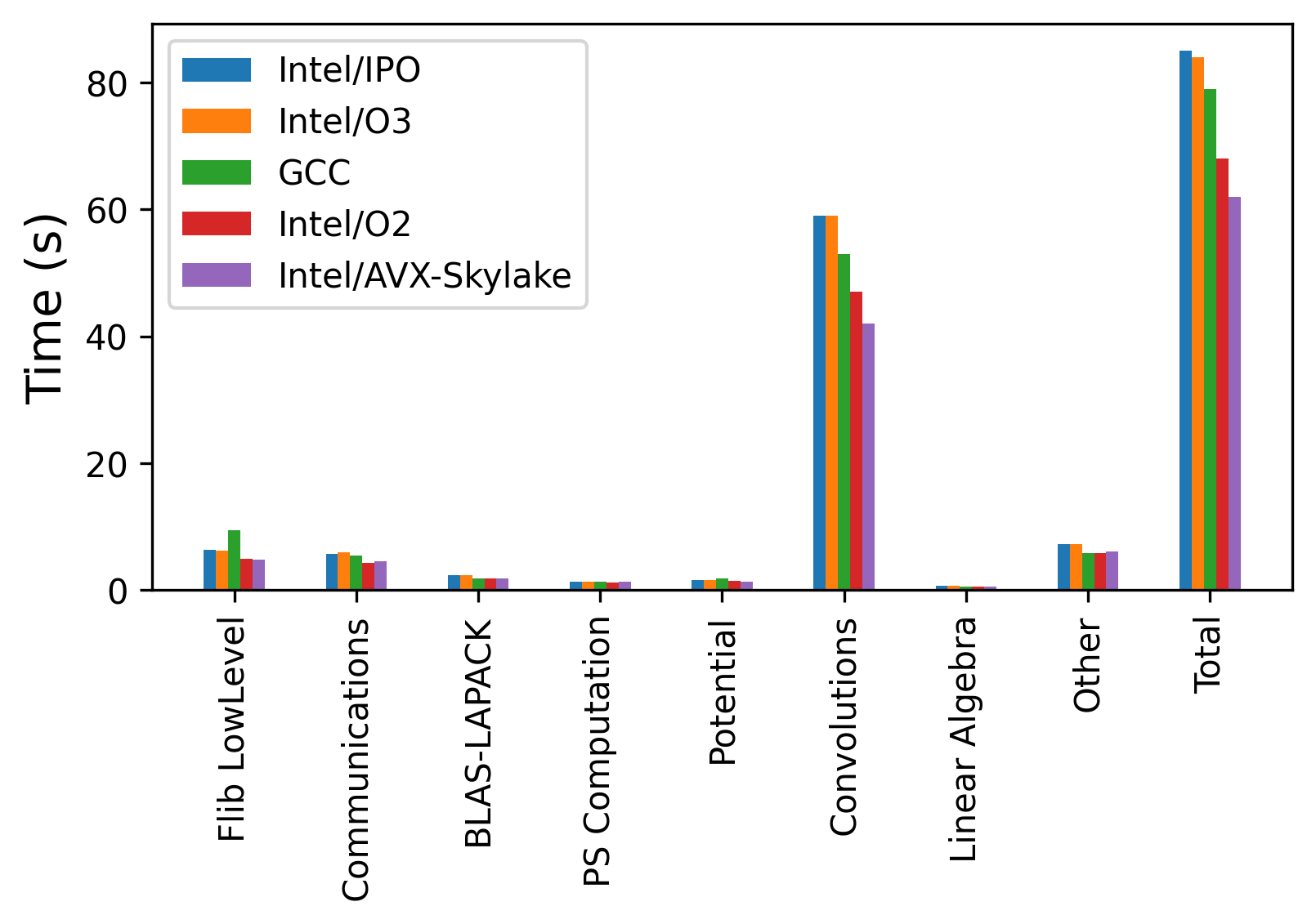}
    \caption{Calculation time for various parts of the BigDFT code built with different compilers and flags. Intel/O2: \lstinline{O2 -qopenmp}; Intel/O3: \lstinline{-O3 -qopenmp}; Intel/AVX-Skylake: \lstinline{-O2 -qopenmp -xSKYLAKE-AVX512}; Intel/IPO: \lstinline{-O2 -qopenmp -ipo}; GCC/O2: \lstinline{-O3 -fopenmp -march=skylake-avx512}.}
    \label{fig:compiler}
\end{figure}

Another example workflow that encapsulates processes performed on both the front end and compute nodes is the tuning of a code through the choice of compiler and flags (List.~\ref{lst:csnip}). We explored five different combinations using a 2CzPN dimer system in the cubic scaling mode of BigDFT (Figure~\ref{fig:compiler}). The code was benchmarked on an Intel Xeon Gold 6152 using 4 MPI processes with 11 threads each. We used version 2021.5.0 of the Intel compiler and 7.3.1 of GCC, with the MKL library for both setups. We found that, for this particular cluster, the difference in architecture between the front end and compute nodes meant that the performance of the convolution routines benefited from explicitly setting the target architecture. Conversely, the inclusion of interprocedural optimization degraded performance. The notebook design relies heavily on cell magics to represent the exploratory process of trying different combinations of flags based on previous results.

\begin{code}
\begin{minted}{python}
# Setup Directory Structure
source = join(expanduser("~"), sdir, "bigdft-suite-1.9.4")
updir = join(expanduser("~"), udir)
build = join(expanduser("~"), bdir)
makedirs(build, exist_ok=True)
makedirs(updir, exist_ok=True)

# Write the buildrc Configuration File.
if upstream:
    with open(join(updir, "buildrc"), "w") as ofile:
        ofile.write(buildrc)
else:
    with open(join(build, "buildrc"), "w") as ofile:
        ofile.write("extra_prefixes=['" + 
                    join(expanduser("~"), udir, "install") + 
                    "']\n")
        ofile.write(buildrc)

# Change to the Build Directory and Compile
if upstream:
    chdir(updir)
    system("python " + join(source, "bundler", "jhbuild.py") 
           + " -f buildrc build upstream-suite")
else:
    chdir(build)
    system("python " + 
           join(source, "Installer.py -y build -a no_upstream"))
\end{minted}
\caption{The function run remotely to compile the BigDFT code.  To build BigDFT is a two step process that involves compiling upstream packages and then the main program. Compilation of the code is one example of the benefit of the arbitrary remote function execution capability of the \remotemanager{} package.}
\label{lst:csnip}
\end{code}

\subsection{Validation of BigDFT: High Throughput}

The BigDFT code is based on a wavelet basis set and pseudopotentials. Thanks to its flexible Poisson solver, this paradigm can be applied to a variety of boundary conditions (periodic, surface, wire, free). This makes BigDFT an unusual code in that it uses pseudopotentials for calculations of molecular systems, which are potentially far outside the training data used to construct these potentials. To validate BigDFT's application to molecular systems, we compared it with the Gaussian based PSI4 code, using the W4-11~\cite{Karton2011} dataset of atomization energies and the def2-QZVP basis set~\cite{Weigend2005}. We employed three different functionals (PBE~\cite{10.1103/PhysRevLett.77.3865}, PBE0~\cite{Adamo1999}, B3LYP~\cite{Stephens1994}), two sets of pseudopotentials (Krack~\cite{Krack2005}, Saha~\cite{saha2017soft}), and compared against the PCSEG basis set series~\cite{Jensen2014}. BigDFT calculations are performed with a grid spacing of 0.37 Bohr for Krack and 0.45 Bohr for the softer Saha NLCC (non-linear core correction) pseudopotentials. PSI4 calculations are performed with the default parameters (spin unrestricted) except that the DFT grid is increased to match Gaussian's UltraFine grid. We removed molecules with sulfur, for which there is no Saha pseudopotential, and \ch{C2} and \ch{ClO2} for which PSI4 had convergence problems with the default settings; in total, there were $3
\times 7\times 136=2856$ calculations to be performed. 

\begin{figure}
    \centering
    \includegraphics[width=1\columnwidth]{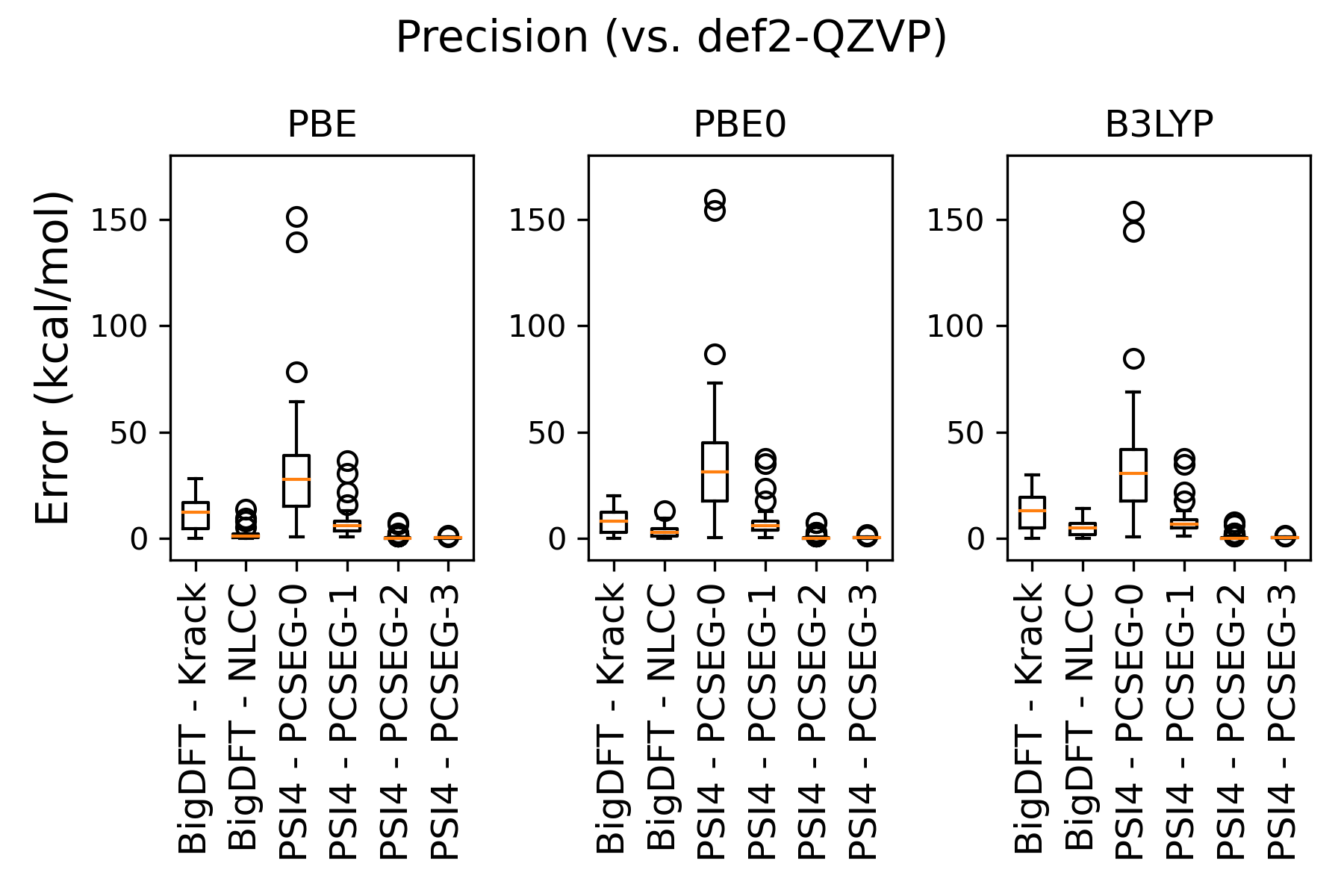}
    \caption{Precision of various calculation methods for the atomization energies of the W4-11 dataset compared to PSI4 calculations with the def2-QZVP basis set. When the NLCC pseudopotentials are employed, the precision is similar to the triplet-$\zeta$ PCSEG-2 basis set. This level of precision appears to be consistent across the different functionals.}
    \label{fig:thermal}
\end{figure}

\begin{code}
\begin{minted}{python}
# Setup Input Parameters
cinp = Inputfile()
cinp.set_xc(functional)

if pp == "Saha":
    cinp.set_psp_nlcc()
    cinp.set_hgrid(0.45)
elif pp == "Krack":
    cinp.set_psp_krack()
    cinp.set_hgrid(0.37)
param_id = str(pp) + "_" + functional

# Iterate
logfiles = {}
for k, v in systems.items():
    # Set Charge and Multiplicity
    cinp2 = deepcopy(cinp)
    cinp2.charge(charges[k])
    if mults[k] > 1:
        cinp2.spin_polarize(mpol = mults[k]-1)
    logfiles[k] = calc.run(sys=v, input=cinp2, name=k,
                           run_dir="bigdft_" + param_id)

return {k: v.energy for k, v in logfiles.items()}
\end{minted}
\caption{The function run remotely to process the W4-11 dataset with BigDFT. Parallelization is done over calculation parameters with each function processing every system.}
\label{lst:therm}
\end{code}

The notebook on Github details how the combination of PyBigDFT and \remotemanager{} can encapsulate this workflow: each combination of code, basis set/pseudopotential, and functional is its own \lstinline{Runner} that computes every structure and returns a dictionary of energy values (List.~\ref{lst:therm}). As expected, the use of NLCC pseudopotentials greatly increases the precision of the calculation, even though the grid spacing can be relaxed. The pseudopotentials are also transferable between functionals, despite the initial training set (we note here the recent, more comprehensive study of Li. et al.~\cite{Li2023}). 

\subsection{Exploration of Excited States with T-CDFT}

\begin{figure}
    \centering
    \includegraphics[width=1\columnwidth]{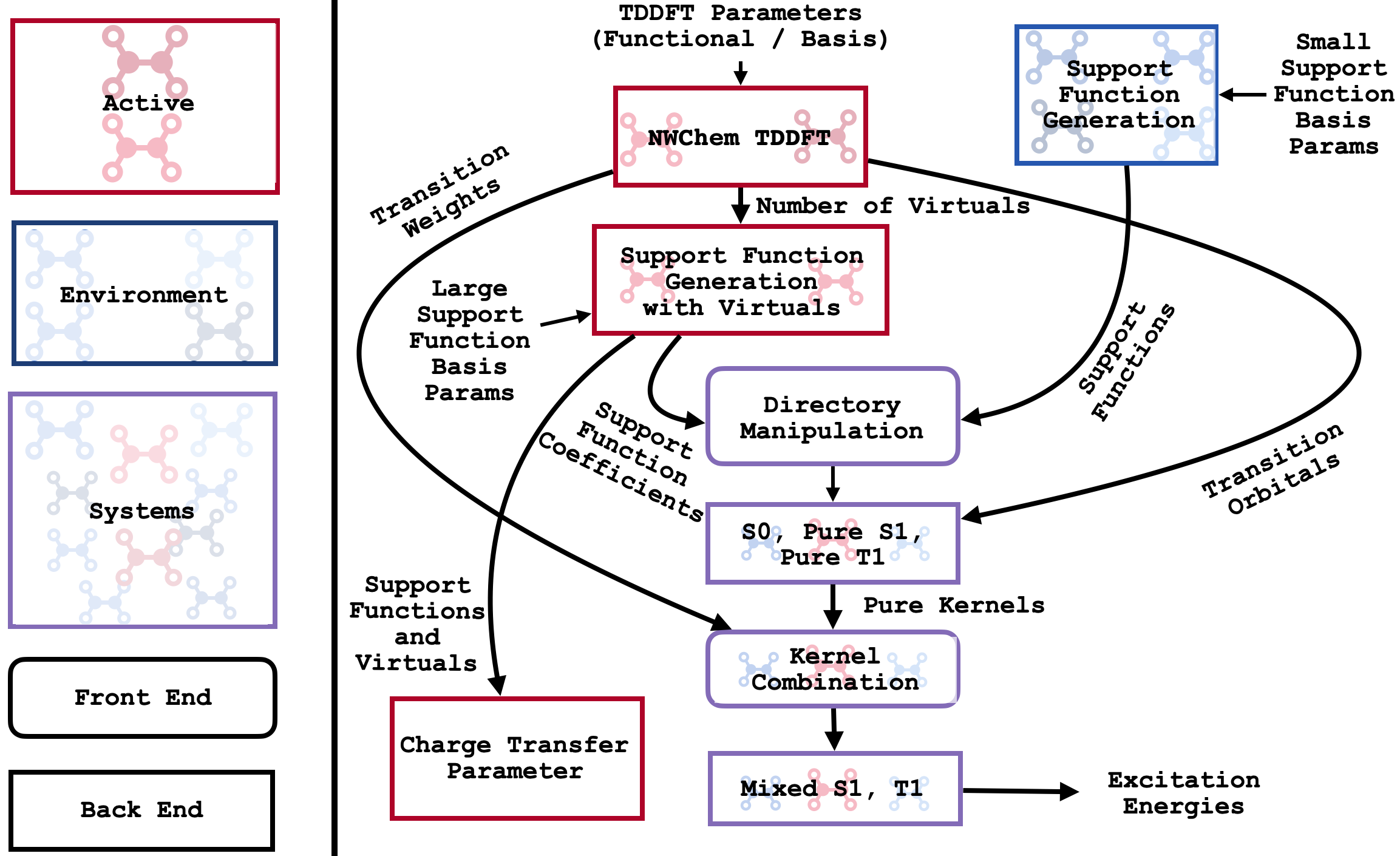}
    \caption{Flow chart of a workflow for calculating excited state singlet and triplet energies of large supramolecular morphologies using Transition-Based Constrained DFT, built on PyBigDFT and \remotemanager. Calculations on active molecules, environment molecules, and the full systems are highlighted in red, blue, and purple, respectively. Calculations run on the back end of the remote machine are in squares while front end calculations have rounded corners. }
    \label{fig:tcdft}
\end{figure}

Recently, we have introduced and implemented the Transition-Based Constrained DFT method for studying excited states~\cite{Stella2022}. The aim of T-CDFT is to provide a method for studying excitations in disordered supramolecular morphologies, which can easily be used in conjunction with BigDFT's molecular fragment approach~\cite{Ratcliff2015,Ratcliff2015b} to reach the required large system sizes. In practice, this requires a workflow that involves many steps, from setting up a system containing ``active'' (i.e.\ those undergoing an excitation) and ``environment'' molecules, to defining the transition to be imposed (e.g.\ based on the outcome of a time-dependent (TD) DFT calculation), to generating basis sets for template gas phase molecules, to calculating excitation energies for pure, e.g.\ HOMO to LUMO, and/or mixed transition constraints on the full systems. For this latter step, it is first necessary to combine the converged density kernels associated with the pure constraints, a process which is run on the front end, as illustrated in List.~\ref{lst:cdft}. Running this kernel on the remote machine avoids the need to copy matrices to one's own machine and allows us to build ``Post-DFT'' functionality on top of an unchanged DFT code (instead of incorporating this mixing into BigDFT's Fortran).  Figure~\ref{fig:tcdft} illustrates a generalised workflow for such a calculation, as well as optional additional post-processing steps, such as characterizing the imposed transitions via the means of a charge transfer parameter.


\begin{code}
\begin{minted}{python}
@RemoteFunction
def combine_density_kernels(kernel_info):
    # Combine the Density Kernels from Calculations with a Pure Constraint
    for p, kernel_dir in enumerate(kernel_info['kernel_dirs']):
        kernel_file = join(kernel_dir, 'density_kernel_sparse.mtx')
        rho_a_p = mmread(kernel_file)

        # Set the Kernel to Zero on the First Step
        if p == 0:
            rho_a = deepcopy(rho_a_p)
            rho_a.data[:] = 0.0

        rho_a += kernel_info['transition_weights'][p] * rho_a_p

    # Write the Kernel for Reading by BigDFT
    mmwrite(join(kernel_info['target_dir'], 'density_kernel_sparse.mtx'), rho_a)
\end{minted}
\caption{A \lstinline{@RemoteFunction} used in the T-CDFT workflow. By decorating this function, it can now be called by any \lstinline{Dataset}. The function takes density kernels generated from T-CDFT calculations with pure constraints, and combines them according to weights taken from the transition breakdown coming from a TDDFT calculation, to be used in a calculation with a mixed constraint. For the generic case of a mixed transition, it is then enough to split the T-CDFT approach into multiple constraints. We can define the energy of the mixed transition by the SCF energy obtained from the density operator $\op \rho^a = \sum_p \mathcal P_p^a \op \rho_p^a $, where $\op \rho_p^a$ is the SCF density obtained from the pure T-CDFT calculation with the constraint $\trace{\dm \op T_p^{(a)}} = 1$.}
\label{lst:cdft}
\end{code}

Since TDDFT in BigDFT can currently only be employed for LDA calculations, the workflow instead uses NWChem~\cite{Apra2020} for the initial step of identifying the transition constraint that will be imposed in later steps, although this step could equally be substituted with another code or approach. Nonetheless, this integration of NWChem into the workflow highlights the benefits of the ability to execute arbitrary Python code with \remotemanager. While other packages provide access to a popular code like NWChem, they may not provide access to all calculation methods (for example, ASE doesn't have settings for TDDFT), or to all the computed properties. For a specific use case though, it is easy to use Python as a glue code to build the appropriate input files and parse the results. The \remotemanager{} class transparently allows the user to run that glue code on a remote machine over a large set of molecules.

\section{Conclusion}

In this Technical Note, we have presented two libraries we have been developing that can enable new kinds of QM calculations on High Performance Computers. The PyBigDFT library is used to set up the system to be studied and to run calculations with the BigDFT code. The \remotemanager{} library offers utilities to asynchronously run these calculations (or any arbitrary Python function) straight from a Python script. These libraries can be deployed alongside Jupyter notebooks to enable a range of activities including exploratory calculations, literate programming, and the building of complex workflows that are applied to large datasets.

An important benefit of the PyBigDFT and \remotemanager{} capability presented here is for distribution of the code. It is now possible to install a BigDFT client version, which contains only PyBigDFT and \remotemanager{}. These packages are significantly easier to install than the full BigDFT suite (both are available on the PyPI repository). The client can be interfaced to a compute cluster with a full version of BigDFT installed once by the system administrators or available using containers like Docker or Singularity.

We have attempted to minimize the requirements needed for the \remotemanager{} library: there is no software that needs to be installed on the remote machine, no daemons or databases to run, and interactions with the queuing systems are deferred to the computer definition. A significant requirement, however, is that it must be possible to login to the machine without a password. The \remotemanager{} library is built on top of SSH, so this can be accomplished using public key authentication or sshpass (when both a key and password is required); for machines with stricter security requirements (such as one-time passwords), \remotemanager{} won't be usable. In principle, there is no requirement for compatibility between the Python implementation running locally (3.7+) and on the remote machine (3.5+), except when using the Dill protocol for serialization. 

In this article, we have focused on literate programming in the context of data analysis, Python programming, and Jupyter notebooks. Another popular tool for literate programming is Emacs Org-mode. Org-mode has already been employed to build software in the field of QM modeling; in particular, we highlight the recent development of packages like Atrip~\cite{Schafer2021, atripsource}, QMCkl~\cite{trexcoeQMCklSource}, and TREXIO~\cite{Posenitskiy2023}. Such projects show the potential for literate programming to lead to full fledged software packages for the QM modeling community. These developments, and those presented here, hint of a future for the scientific community where new theoretical methods and computational schemes are shared in ways that bring together scientific narratives and concrete implementations.

\section*{Acknowledgements}

Computations were performed using resources at the Research Center for Computational Science, Okazaki, Japan (Project: 23-IMS-C029). Calculations were also performed using the Hokusai supercomputer system at RIKEN (Project ID: Q23460). This work was supported by MEXT as ``Program for Promoting Research on the Supercomputer Fugaku'' (Realization of innovative light energy conversion materials utilizing the supercomputer Fugaku, Grant Number JPMXP1020210317). This work also used the ARCHER2 UK National Supercomputing Service (https://www.archer2.ac.uk), and the computational facilities of the Advanced Computing Research Centre, University of Bristol (http://www.bris.ac.uk/acrc/). LG, NT, and WD acknowledge the joint CEA--RIKEN collaborative action. LER and MS acknowledge support from an EPSRC Early Career Research Fellowship (EP/P033253/1). 
\section*{References}

\bibliographystyle{unsrt}
\bibliography{./bibliography/computational,./bibliography/theoretical,./bibliography/bigdft,./bibliography/reviews,./bibliography/applications,./bibliography/software}

\end{document}